\documentclass[12pt,a4paper,twocolumn]{narms}
\usepackage{graphics}

\title {An MAS-Based ETL Approach for Complex Data}

\author{\vspace{0.25cm}O. Boussaid, F. Bentayeb,  J. Darmont\\
\small{ERIC/BDD, Universit{\'e}  Lumi{\`e}re Lyon 2}\\
\small{5 avenue Pierre-Mend\`es-France}\\
\small{Bron Cedex}\\
\small{FRANCE}\\
\small{\emph{E-mail: boussaid@univ-lyon2.fr, bentayeb@eric.univ-lyon2.fr,jerome.darmont@univ-lyon2.fr}}}

\date{}

\input epsf

\begin{document}

\setcounter{page}{0} \pagenumbering{arabic}

\maketitle

\begin{abstract}

\small {{\bf Abstract:} In a data warehousing process, the phase of
data integration is crucial. Many methods for data integration
have been published in the literature.  However, with the development of the
Internet, the availability of various types of data (images,
texts, sounds, videos, databases...) has increased, and structuring
such data is a difficult task. We name these data, which may be structured or
unstructured, "complex data". In this paper, we  propose a
new approach for complex data integration, based on a Multi-Agent
System (MAS), in association to a data warehousing approach. Our
objective is to take advantage of the MAS
to perform the integration phase for complex data. We indeed
consider the different tasks of the data integration process as
services offered by agents. To validate this approach, we have
actually developped an MAS for complex data integration.

}

\end{abstract}

\section{Introduction}

%

The data warehousing and OLAP (On-Line Analytical Processing)
technologies \cite{Inm96,Kim96} are now considered mature in
management applications, especially when data are numerical. With
the development of the Internet, the availability of various types
of data (images, texts, sounds, videos, databases...) has
increased. These data, which may be structured or unstructured, are
called "complex data". Structuring and exploiting these data
is a difficult task and requires the use of efficient techniques and
powerful tools to facilitate their integration into a data
warehouse. It actually consists in Extracting and Transforming complex
data before they are Loaded in the data warehouse (ETL process).

In this paper, we propose a new approach for complex data
integration, based on a Multi-Agent System (MAS). Our approach
consists in physically integrating complex data into a relational
database (an ODS -- Operating Data Storage) that we consider as a
buffer ahead of the data warehouse. We are then interested in
extracting, transforming and loading complex data into the ODS.

The aim of this paper is to take advantages of Multi-Agent
Systems that are intelligent programs, composed of a set of
agents, each one offering a set of services, to perform 
complex data integration. We can indeed assimilate the different
tasks  of the integration process, which is technically difficult,
to services carried out by agents.

\noindent \emph{Data extraction}: This task is performed by an
agent in charge of extracting data characteristics from complex data.
The obtained characteristics are then transmitted to  an agent
responsible for data structuring.

\noindent \emph{Data structuring}: To perform this task, an
agent deals with the organization of data
according to a well-defined data model.
 Then, this model is transmitted to an agent responsible for data storage.

\noindent \emph{Data storage}: This task is performed by an agent
that
feeds the database with the source data, using the model supplied by the data structuring agent.\\

In order to validate this approach, we have designed an MAS for
complex data integration. This system is composed of a set of
intelligent agents offering the different services that are
necessary to achieve the integration process of complex data. It is
based on an evolutionary architecture that offers a great
flexibility. Our system indeed allows to update the existing
services or to add/create new agents.

The remainder of this paper is organized as follows. Section~\ref{StateArt}
presents a state of the art regarding data integration
approaches and agent technology. In Section~\ref{ETL}, we present
the issue of complex data integration and our  approach.  We explain
the advantages of MASs in Section~\ref{justification} and show why
they are  adapted to carry out this approach via our proposed
architecture. Finally, we conclude this paper and present
research perspectives in Section~\ref{conclusion}.

\section{State of the art}\label{StateArt}
%
%

We present in this section an overview of the techniques our
proposal relies on, namely those regarding data integration,
the ETL (Extracting, Transforming, and Loading) process, and Multi-Agent Systems.

\subsection{Data integration}

Nowadays, two main and opposed approaches are used to perform data
integration over the Web.

In the mediator-based approach \cite{MCR02}, the different data
remain located at their original sources. The user is provided an
abstract view of the data, which represents distributed and
heterogeneous data as if they were stored in a centralized and
homogeneous system. The user's queries are executed through a
mediator-wrapper system \cite{GLR00}. A mediator reformulates
queries according to the content of the various accessible data
sources. A wrapper is data source-specific, and extracts the
selected data from the target source. The major interest of this
approach is its flexibility, since mediators are able to
reformulate and/or approximate queries to better satisfy the user.
However, when the data sources are updated, modified data are
lost, which is not pertinent in a decision support context where
historicity is important.

On the opposite, in the data warehouse
approach~\cite{Inm96,Kim96}, all the data from the various data
sources are centralized in a new database, the data warehouse. The
multidimensional data model of a data warehouse is
analysis-oriented: data represent indicators (measures) that can
be observed according to axes of analysis (dimensions). A data
warehouse actually characterizes and is optimized for one given
analysis context. In a data warehouse context, data integration
corresponds to the ETL process that accesses to, cleans and
transforms the heterogeneous data before they are loaded in the
data warehouse. This approach supports the dating of data and is
tailored for analysis. However, refreshing a data warehouse is a
complex and time-consuming task that implies running a whole ETL
process again each time an update is required.

\subsection{ETL process}

The classical ETL process, as its name hints, proceeds in three
steps~\cite{Kim96}. The first \emph{extraction} phase includes
understanding and reading the data source, and copying the
necessary data in a buffer called the preparation zone. Then, the
second \emph{transformation} phase proceeds in several successive
steps: clean the data from the preparation zone (syntactic errors,
domain conflicts, etc.); discard some useless data fields; combine
the data sources (by matching keys, for instance); create new keys
for dimensional records to avoid using keys that are specific to
data sources; and build aggregates to optimize the more frequent
queries. In this phase, metadata are essential to store the
transformation rules and various correspondences. Eventually, the
third \emph{loading} phase stores the prepared data into
multidimensional structures (data warehouse or data marts). It
also usually includes an indexing phase to optimize later
accesses.

\subsection{Multi-Agent Systems}

An agent software is a classical program that is  qualified as
"intelligent". Intelligent agents are used in many fields such
as networks, on-board technologies, human learning... An
intelligent agent is supposed to have the following intrinsic
characteristics: \emph{intuitive} -- it must be able to take
initiatives and to complete the actions that are assigned to it;
\emph{reactive} -- it must be aware of its environment
and act in consequence; \emph{sociable} -- it must be able to
communicate with other agents and/or users~\cite{klus01}. Moreover,
agents may be mobile and can independently move through an
acceptor network in order to perform various tasks. 

A Multi-Agent
System  designates a collection of actors that communicate with
each other~\cite{syk96}. Each actor is able to offer specific
services and has a well-defined goal. This introduces the concept
of service: each agent is able to perform several tasks, in an
autonomous way, and communicates the results to a receiving actor
(human or software). The MAS must respect the programming
standards defined by the FIPA (Foundation for Intelligent Physical
Agents)~\cite{fip02}.

\section{MAS-based approach for complex data ETL}\label{ETL}


\subsection{Complex data integration approach}

Data integration corresponds to the ETL phase in the data warehousing
process. To achieve the integration of complex data, the traditional
ETL approach is however not adapted. We present in this paper our
approach to accomplish the extracting, transforming, and loading
process on complex data in an original way.

\begin{figure}[hbt]
\begin{center}
\epsfxsize=7cm \centerline{\epsffile{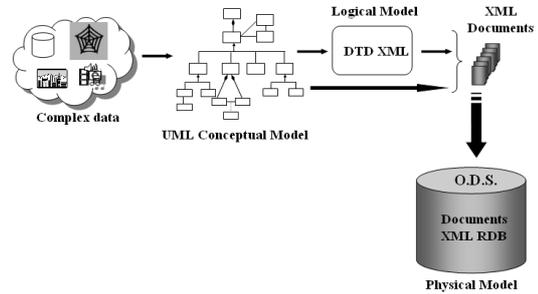}} \caption{
Classical modelling process for complex data
integration}\label{umlxml}
\end{center}
\end{figure}

In order to integrate complex data captured from the Web, for instance, into a decision
support database such as a data warehouse, we have proposed a full modelling
process (Figure~\ref{umlxml}). We first designed a conceptual UML
model for a complex object representing a superclass of all the
types of complex data we consider (text, multimedia documents,
relational views from databases)~\cite{DBB02}. The UML conceptual
model is then directly translated into an XML schema (DTD or
XML-Schema), which we view as a logical model. The last step in
our (classical) modelling process is the production of a physical
model in the form of XML documents that are stored in relational
database. We consider this database as an ODS (Operational Data
Storage), which is a data repository that is typically used in a
traditional ETL process before the data warehouse proper is
constituted. However, note that our objective is not only to store data, but also to
truly prepare them for analysis.

\subsection{MAS-based prototype}

\begin{figure}[hbt]
\begin{center}
\epsfxsize=7cm \centerline{\epsffile{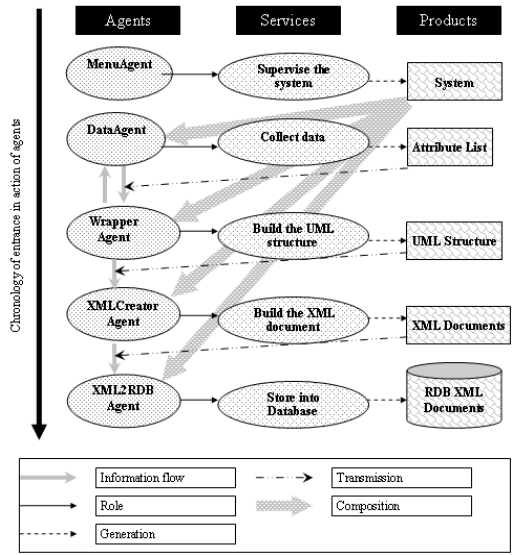}} \caption{MAS-based ETL architecture for 
complex data integration}\label{system}
\end{center}
\end{figure}

The integration of complex data is more difficult than a
classical ETL process. This technically difficult
integration process requires a succession of tasks that we assimilate to
services that may be carried out by agents. To effectively achieve this goal, 
we have designed an MAS-based prototype. Its
architecture is presented in Figure~\ref{system}. It is based on a platform of
generic agents. We have instantiated five agents offering
services that allow the integration of complex data. The purpose of
this collection of agents is to perform several tasks. Each agent
is able to offer specific services and has a well-defined goal.

The first main agent created in our prototype, 
\emph{MenuAgent}, pilots the system, supervises agent
migrations, and indexes the accessible sites from the platform.
Some others default pilot agents help in the management of the
agents and provide an interface for the agent development
platform.

The essential of the integration process is achieved through
services about collecting, structuring, generating and storing
data, provided by the remaining agents we present in the
next section.

To develop our prototype, we have built a platform  using JADE
version 2.61~\cite{jad02} and the Java language~\cite{jav02}, which is
portable across agent
programming platforms. The prototype is freely available on-line~\cite{SMAIDoC-Download}.

\subsection{Complex data ETL}

\subsubsection{Extracting}

Recall that our modelling approach corresponds to the complex data
integration process. The conceptual level helps the user 
selecting the data and establishing its analysis goals. The
\emph{Extraction} phase is thus carried out by the \emph{DataAgent} agent that
collects the data concerning the documents. This task  consists in
extracting the attributes of the complex object that has been
selected by the user. A particular treatment is applied, depending
on the subdocument class (image, sound, etc.), since each subdocument class bears different attributes.
The \emph{DataAgent} agent uses three ways to extract the actual data: (1)
it communicates with the user through graphical interfaces,
allowing a manual capture of data; (2) it uses standard Java
methods and packages; (3) it uses other ad-hoc automatic extraction
algorithms~\cite{BBDRZ02}. Our objective is to progressively
reduce the number of manually-captured attributes and to add new
attributes that would be useful for later analysis and that could
be obtained with data mining techniques. This work is completed
by the \emph{WrapperAgent} agent that instantiates the UML structure based
on the data supplied by the \emph{DataAgent} agent.

\subsubsection{Transforming}

The logical level coincides with the \emph{Transforming} phase.
Our UML conceptual model is directly translated by
the \emph{XMLCreator} agent into an XML schema (DTD or
XML-Schema), which we view as a logical model. XML is the format
of choice for both storing and describing the data. The schema
indeed represents the metadata. XML is also very interesting
because of its flexibility and extensibility, while allowing
straight mapping into a conventional database if strong
structuring and retrieval efficiency are needed for
analysis purposes.

\subsubsection{Loading}

 The last level  in our modelling process corresponds
to the \emph{Loading} phase. It consists in the production of a
physical model in the form of XML documents and their loading into
a relational database. This is achieved by the \emph{XMLCreator} and
\emph{XML2RDBAgent} agents. The principle of the \emph{XMLCreator} agent's service is
to parse the XML schema recursively, fetching the elements it
describes, and to write them into the output XML document, along
with the associated values extracted from the original data, on
the fly. Missing values are currently treated by inserting an
empty element, but strategies could be devised to solve this
problem, either by prompting the user or automatically. The XML
documents obtained with the help of the \emph{XMLCreator} agent are mapped
into a relational database by the \emph{XML2RDBAgent} agent. It operates in
two steps. First, a DTD parser exploits our logical model (XML
schema) to build a relational schema, i.e., a set of tables in
which any valid XML document (regarding our DTD) can be mapped. To
achieve this goal, we mainly used the techniques proposed by
\cite{Der00,Kap00}. Note that our DTD parser is a generic tool: it
can operate on any DTD. It takes into account all the XML element
types we need, e.g., elements with +, *, or ? multiplicity,
element lists, selections, etc. The last and easiest step consists
in loading the valid XML documents into the previously build
relational structure.

\section{Justification}\label{justification}

The variety of data types (images, texts, sounds, videos,
databases...) increases the complexity of data. It is thus necessary to
structure them in an "un-classical" way. Because data are complex,
they necessitate more information. Furthermore, it is important to
consider this information and to represent it in the form of metadata. 
Then, the choice of the XML formalism is fully justified.
Since our proposal is based on a classical modelling process, it
allows the user to determine what are his/her analysis objectives,
to select how to represent the data and how to store them into a
database. It constitutes a whole process permitting to carry out
the integration of complex data. This is also the objective of
the ETL process.

Our proposed process necessitates several tasks that must be performed,
repetitively. These tasks are not necessarily sequential, and
are assimilated to services offered by well-defined agents in a
system intended to achieve such an integration process. With 
this goalin mind, we have developed a MAS-based prototype that
is based upon a flexible and evolutive architecture on which we
can updated services, and even create new agents to consider data
refreshing, analysis and so on.

%

 \section{Conclusion and Perspectives}\label{conclusion}

In this paper,  we have proposed a new approach for  complex data
integration based on both the data warehouse technology and
multi-agent systems. This approach is based on a flexible and
evolutive architecture on which we can add, remove or modify
services, and even create new agents. We then developped a
MAS-based prototype that allows this integration with respect to
the following three steps of the ETL process. Two agents named
 \emph{DataAgent} and \emph{WrapperAgent}, respectively, 
model the input complex data into UML classes. The
\emph{XMLCreator} agent translates UML classes into XML documents
that are mapped in a relational database by the
\emph{XML2RDBAgent} agent. Moreover, note that the different agents that
compose our system are mobile and that the services they propose
coincide with the  ETL asks of the data warehousing process.

 We plan to extend the services offered by our MAS-based
prototype, especially for extracting data from their sources and
analyzing  them. For example, the \emph{DataAgent} agent could converse
with on-line search engines and exploit their answers. On the other
hand, we could also create new agents in charge of modelling data 
multidimensionally in order to apply analysis methods such as OLAP or data
mining.

\bibliographystyle{alpha}
\small{\bibliography{My_Bibliography}}

\end{document}